\documentclass[journal]{IEEEtran}

\usepackage{cite}
\usepackage{epstopdf}
\usepackage{bm}
\usepackage{amsmath}
\usepackage{amssymb}
\usepackage{pifont}
\usepackage{xtab}
\usepackage{mathrsfs}
\usepackage{subeqnarray}
\usepackage{cases}
\usepackage{booktabs}
\usepackage{multirow}
\usepackage{mathtools}
\usepackage{enumitem}
\usepackage{empheq}
\usepackage{caption}
\usepackage{subcaption}
\allowdisplaybreaks
\usepackage{xcolor,cite,etoolbox}

\makeatletter 
\pretocmd\@bibitem{\color{black}\csname keycolor#1\endcsname}{}{\fail}
\newcommand\citecolor[1]{\@namedef{keycolor#1}{\color{blue}}}
\makeatother
\IEEEoverridecommandlockouts

\newcommand{\newstuff}[1]{\color{black}{#1}\color{black}}

\begin{document}

\title{\huge Practical Challenges in Real-time Demand Response\thanks{This work was supported by ARPA-E award No. DE-AR0000702 and NU’s Finite Earth Initiative (funded by Leslie and Mac McQuown). FLEXLAB’s work was supported by DOE award No. DE-AC02-05CH11231.}}
\author{
\IEEEauthorblockN{Chao Duan, Guna Bharati, Pratyush Chakraborty, Bo Chen, Takashi Nishikawa, Adilson E. Motter
\vspace{-0.3cm}
\thanks{C. Duan, T. Nishikawa, and A. E. Motter are with Northwestern University. G. Bharati is with OPAL-RT Corporation. Pratyush Chakraborty is with BITS Pilani. Bo Chen is with Argonne National Laboratory.}%
%\thanks{G. Bharati is with OPAL-RT Corporation.}%
%\thanks{P. Chakraborty was with the Department of Physics and Astronomy, Northwestern University, Evanston, IL 60208 USA. He is now with the Department of Electrical and Electronics Engineering, Birla Institute of Technology \& Science (BITS), Pilani, India.}
%\thanks{Bo Chen is with Argonne National Laboratory.}%
}
% <-this % stops a space

\vspace{-0.7cm}
}

\maketitle

\begin{abstract}
We report on a real-time demand response experiment with 100 controllable devices. The experiment reveals several key challenges in the deployment of a real-time demand response program, including time delays, \newstuff{uncertainties}, characterization errors, multiple timescales, and nonlinearity, which have been largely ignored in previous studies. To resolve these practical issues, we develop and implement a two-level multi-loop control structure integrating feed-forward proportional-integral controllers and optimization solvers in closed loops, which eliminates steady-state errors and improves the dynamical performance of the overall building response. The proposed methods are validated by Hardware-in-the-Loop (HiL) tests.
\end{abstract}


% \IEEEpeerreviewmaketitle

\vspace{-0.3cm}
\section{Introduction}

\IEEEPARstart{D}EMAND response (DR) gives more flexibility to modern smart grids to compensate for the variability of renewable sources. Different mechanisms \cite{samad2016automated} have been designed to incentivize DR for providing various forms of ancillary services \cite{ma2013demand}, e.g., load shifting and frequency regulation, to support grid operation. In a DR program, diverse and distributed loads in commercial or residential buildings need to be controlled and aggregated to achieve grid-scale effects. In recent years, sophisticated mathematical models and optimization methods have been developed to optimally schedule smart appliances in buildings to achieve desired overall response while respecting the comfort of the users \cite{zhao2013optimal,kim2013scalable,pallonetto2019demand}. These models and methods are built on the assumption that every controllable device will closely follow the control command with its power consumption matching the nominal value. \newstuff{It has been recognized in \cite{torriti2010demand,nolan2015challenges,yassine2016implementation} that DR programs face many practical challenges including economic feasibility, security, user acceptance, and privacy issues. Those insights were mostly gained from conceptual and model analyses and focused on economic and policy barriers. Little field test experience has been reported to expose the extent of engineering challenges.}

\begin{figure}[t]
	\includegraphics[width=3.6in]{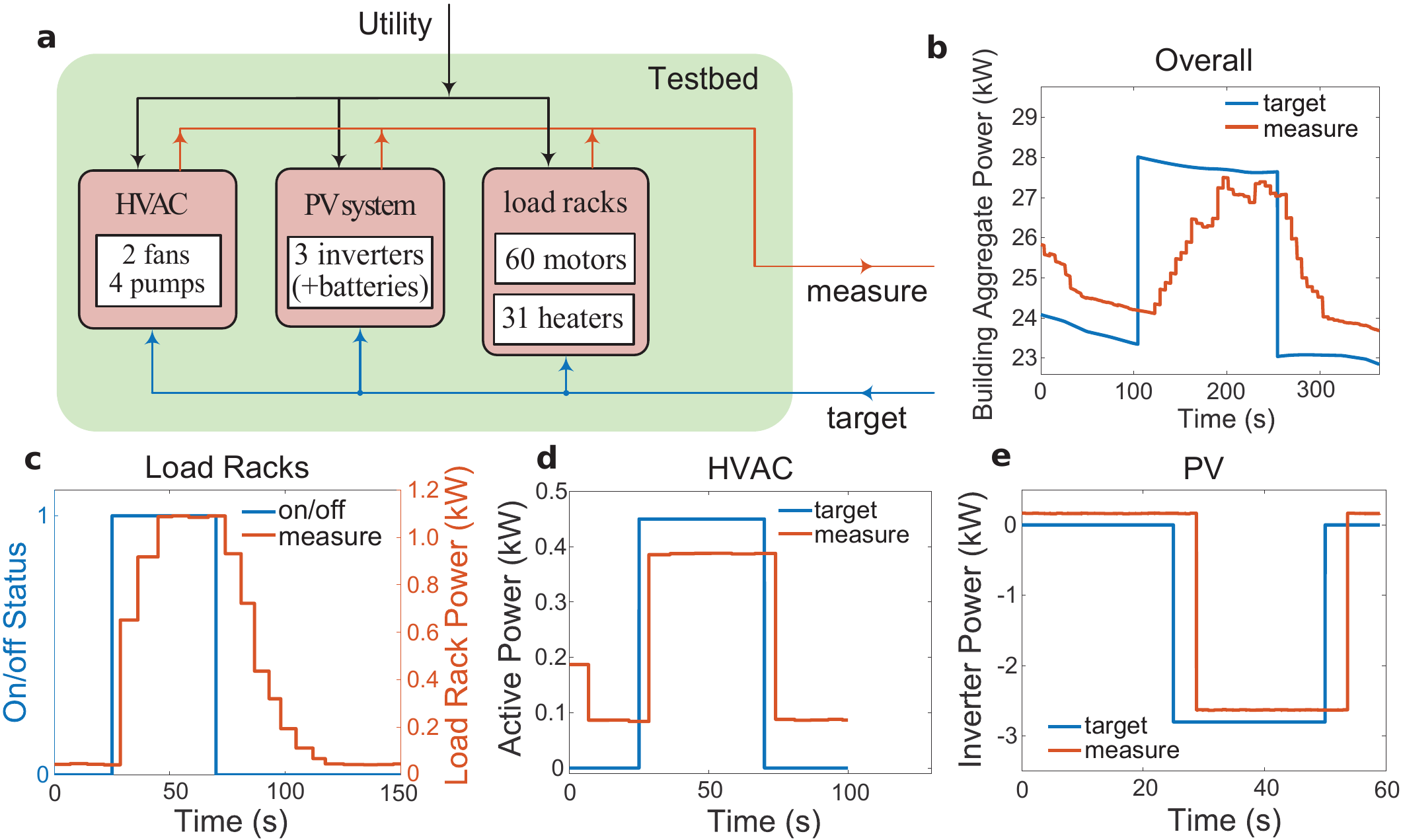}
	\caption{(a) Testbed configuration and (b-e) open-loop test results.}
	\label{LoadRack_Ind}
	\vspace{-0.6cm}
\end{figure}

For DR that aims at improving grid stability (e.g., as in \cite{delavari2018sparse}), the loads are assumed to respond fast enough to participate in the grid dynamics \newstuff{(at the timescale of seconds for frequency regulation)}. \newstuff{Even for DR that acts only on slower timescales, the loads are assumed to reach the set power in the steady state. However, in both cases, the assumptions can \newstuff{be violated substantially in } reality, rendering these DR methods impractical. } To uncover practical challenges, we perform HiL experiments at FLEXLAB, the integrated building and grid testbed at Lawrence Berkeley National Laboratory. The testbed is illustrated in Fig. \ref{LoadRack_Ind}a, in which 100 controllable devices are to be controlled to achieve an aggregate building power target set by an OPAL-RT simulator based on real-time device power feedback. Among these devices, the power of the \newstuff{heating, ventilation, and air conditioning (HVAC) } system (4 pumps and 2 fans) and of the 3 inverters of the photovoltaic (PV) system can be continuously adjusted. For the 31 heaters and 60 blower motors, only their on/off status can be adjusted. We first implement and test an open-loop optimization scheduler that assigns control command to each device based on its prior characterization and specification. This control architecture is assumed in most papers on demand response \cite{zhao2013optimal,kim2013scalable,pallonetto2019demand,delavari2018sparse}. The test results are shown in Figs. \ref{LoadRack_Ind}b-\ref{LoadRack_Ind}e. The overall response of the aggregate building power is very slow \newstuff{(the initial response time is about 4.5 s) } and the desired steady state is never reached (\newstuff{i.e., the ramp time is infinity}). The reasons for \newstuff{the slow dynamical responses and the large steady-state errors } are as follows. First, many devices and the associated communication channels are subject to significant time delays, which vary substantially from device to device (in the range of 1 s to 8 s) \newstuff{and have uncertainties}. Second, many devices \newstuff{have parameter uncertainties } and characterization errors, meaning that their steady-state power consumptions differ significantly from their nominal values. For example, the power consumption of HVAC heavily depends on the indoor and outdoor temperature and the relation is difficult to characterize \emph{a priori}. Third, we have a mixture of continuously and discretely adjustable devices with widely different response times, making it a hybrid multi-timescale system. Fourth, aggregating power consumptions across multiple devices is nonlinear. For example, the power consumption of a blower is 1.5 kW when it is turned on alone, but the average power consumption reduces to 1.1 kW when all the blowers are turned on. Therefore, the impacts of time delays, \newstuff{uncertainties}, characterization errors, multiple timescales, and nonlinearity must be addressed for DR deployment.

%  Control parameters must be designed respecting different time delays.  This causes a significant challenge for the optimal scheduling of devices. Therefore, feedback controllers must be implemented in addition to the optimal scheduling algorithm to eliminate steady-state errors. Third, we may have a mixture of continuously and discretely adjustable devices with widely varying response times, requiring the optimization algorithm and control strategy to work with the hybrid multi-time scale system. Further research may include decentralization of the transmission and distribution control and a systematic approach to deal with time delays, characterization errors, multi-time scales, and discreteness of building-level control involving a large number of controllable devices and complex human behavior.

% Non-Linearity in power consumption was identified in the load rack blowers because of the voltage change. When all the blowers in a load rack were run individually, the sum of the power consumption of the blowers was around 1.5kW. In contrast, when all the blowers were turned ON, the power consumption reduced to 1.1kW. Such non-linearity can result in wrong estimates of device schedules in the demand response program.

\hspace{0.2cm}

% \begin{figure}[ht!]
% 	\centering
% 	\includegraphics[width=2.8in]{HVAC_Ind.pdf}
% 	\caption{Aggregate building power under conventional single loop control} \label{HVAC_Ind}
% \end{figure}

% \begin{figure}[ht!]
% 	\centering
% 	\includegraphics[width=2.8in]{Inverter_Ind.pdf}
% 	\caption{Aggregate building power under conventional single loop control} \label{Inverter_Ind}
% \end{figure}

\vspace{-0.65cm}
\section{A Two-level Multi-loop Control Structure}
\label{hiercon}

% \begin{figure}[h!]
% 	\centering
% 	\includegraphics[width=3.0in]{SimplifiedDiagram.pdf}
% 	\caption{Simplified Control Loops} \label{simplediagram}
% \end{figure}

\begin{figure}[t]
	\includegraphics[width=3.4in]{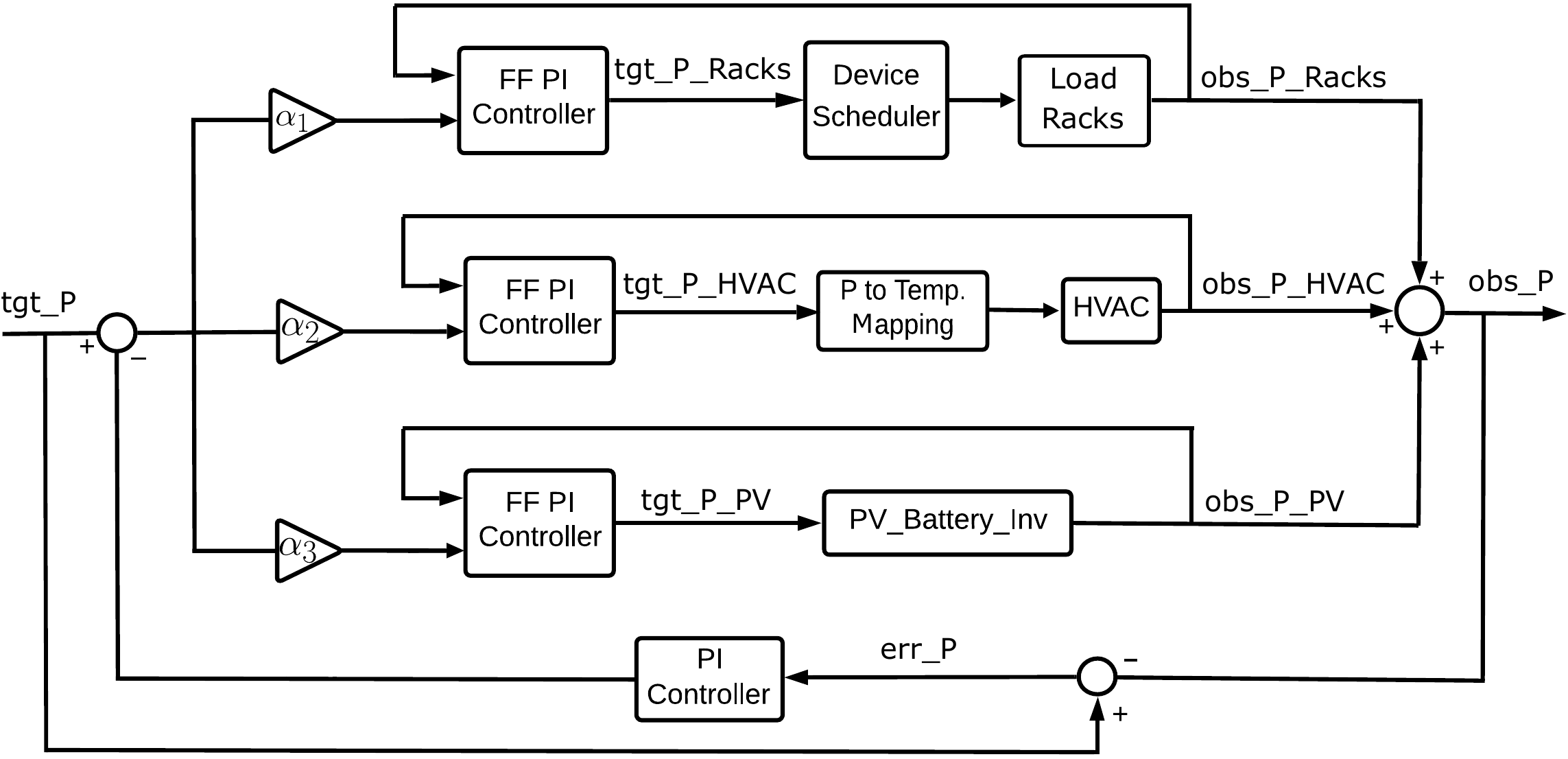}
	\caption{Overall control diagram of the proposed approach.} \label{overall}
	\vspace{-0.6cm}   
\end{figure}

Considering the aforementioned practical challenges, we propose the two-level control structure shown in Fig. \ref{overall} to achieve the desirable performance. At the lower level, different types of devices have different closed-loop time delays and response times. We divide all the devices into three groups, and for each group we implement a feed-forward (FF) proportional–integral (PI) controller to regulate the group-level aggregate power output. Thus, each FF PI controller can be tuned individually according to the dynamical features of the corresponding group. 
% The first group contains the on-off devices including small heaters and motors. The feed-forward PI generates a dynamic power target for the on-off device scheduler which decides the on/off status of the devices. The second group includes the whole HVAC system. The HVAC usually has its own control system which accepts a target temperature as control input, therefore we use a characterization mapping to transform our pwoer target into a temperature value sent to the HVAC system. The third group is the PV/battery/inverter system for which we directly send a power target to the inverter. 
On top of the three parallel FF PI control loops, we implement an upper-level PI controller to regulate the aggregate building power. The output of this classical PI controller is then distributed to the three groups according to a set of participation factors.

\begin{figure}[t]
	\centering
\includegraphics[width=3.2in]{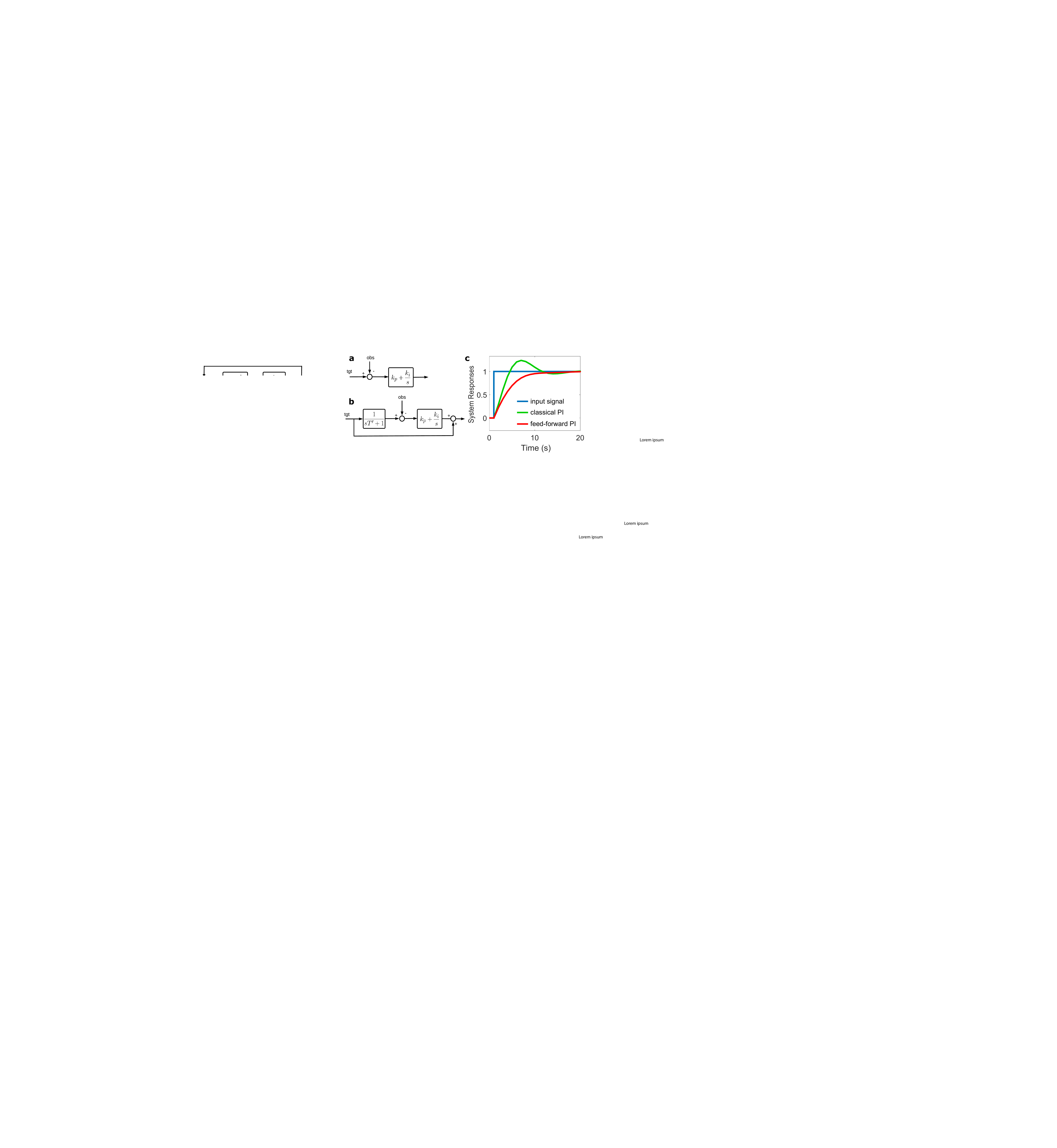}
	\caption{(a) Classical vs. (b) feed-forward PI control. (c) Resulting
responses.} 
\label{purePI}
	\vspace{-0.6cm}   
\end{figure}

\textit{Feed-forward PI control.} The structure of the FF PI controller is shown in Fig. \ref{purePI}b, and it is instructive to compare it with the classical PI (shown in Fig. \ref{purePI}a). Both controllers achieve zero control error in the steady state. The advantage of the FF PI controller is that, when choosing a time constant $T'$ close to the plant time constant $T$ (assuming that the plant can be represented by a first-order transfer function $G(s) =    \frac{h}{sT+1}$), the closed-loop zeros would be very close to the closed-loop poles. Therefore, due to the zero-pole canceling effect, the overall response will resemble a first-order transfer function with steady-state gain of 1. In this way, the FF PI controller corrects the steady-state characterization error of the system without significantly changing the dynamical behavior. A comparison of the responses of systems controlled by the two controllers is presented in Fig. \ref{purePI}c, where $T=5,\ h=1.2,\ k_p=1,\ k_i=1$, and $T'=3.5$. 

% Assume that 1) the dynamics of the optimization schedulers and time delays are neglected; 2) the discreteness of device power is neglected; 3) with the feed forward PI controller, the controlled devices are represented by first-order transfer function $G_i = \frac{1}{sT_i +1}$, $i=1,2,3$. The closed-loop response under input signal $u(s)$ is given by
% \begin{equation}\label{wholeresp}
% \begin{aligned}
%     y =  u  +\frac{G_1u_1+G_2u_2+G_3u_3-u}{1+(\alpha_1G_1+\alpha_2G_2+\alpha_3G_3)(k_P+{k_I}/{s})}
% \end{aligned}
% \end{equation}
% where $u=u_1+u_2+u_3$ and $y=y_1+y_2+y_3$. 
% The individual output for $i=1,2,3$ writes
% \begin{equation}\label{indivitualresp}
%     y_i= G_i \left( u_i + \alpha_i \frac{u-G_1u_1-G_2u_2-G_3u_3}{s/(sk_P+k_I)+\alpha_1G_1+\alpha_2G_2+\alpha_3G_3} \right)
% \end{equation}
\textit{Basic features of the control structure.} The proposed two-level control structure possesses the following desirable features. First, it has \emph{zero steady-state error}. The steady-state response is independent of the PI controller parameters, and it is determined only by the participation factors and device scheduler, i.e., the control scheme achieves the control target with no error in the steady state. Second, it enables \emph{the fastest possible transient response}. During the transient, each type of device respond independently to the overall control error of the system. In this way, faster devices will kick in first to achieve the overall target even if they are not assigned to do so, and the slow devices will follow to take the assigned steady-state power. Therefore, the overall system will have the fastest possible initial response and also the largest possible steady-state capacity. Third, it has \emph{fault tolerance}. When one or more devices fail and cannot respond to the assigned target power, the rest will share the deficit power according to the participation factors. 

\textit{Device scheduler.} Given the target $\hat{P}(t)$ at time $t$, the device scheduler decides the status of on/off loads ($\mathbb{K}_t$) and the power of continuously adjustable loads ($\mathbb{Z}_t$) by solving the following mixed-integer linear program:
\begin{subequations}\label{BLoptmodel}
\begin{align}
    \max_{\bm{u}(t), \bm{v}(t)} \sum _{i \in \mathbb{K}_t} u_{i}(t) \cdot f_{i} + \sum _{j \in \mathbb{Z}_t } v_{j}(t),
\end{align}
\begin{empheq}[left =  \mathrm{s.t.}\empheqlbrace\,]{align}
 &  \hat{q}(t) +\sum _{i \in \mathbb{K}_{t}}u_{i}(t) \cdot f_{i}  +\sum _{j \in \mathbb{Z}_t}v_{j}(t)  \le \hat{P}(t), \\
&   v_{j}^{\min}(t) \le v_{j}(t) \le v_{j}^{\max}(t),\  u_{i}(t) \in \left\{0,1\right\}.
\end{empheq}
\end{subequations}
Here, $\hat{q}(t)$ is the total power consumption of the uncontrollable devices, $u_{i}(t)$ is the binary variable representing the status of the $i$th on/off load, $f_{i}$ is the power consumption of the $i$th on/off load when it is on, $v_{j}(t)$ is the power consumed by the $j$th continuously adjustable load with $v_{j}^{\min}(t)$ and $v_{j}^{\max}(t) $ being its lower and upper limits, respectively. \newstuff{For scalability, one can organize the loads into several groups, each with its own independent device scheduler.}

\textit{Stability analysis with time delays.} Time delays may significantly degrade the dynamical performance or even destabilize the system. Therefore, the control design must account for time delays to ensure acceptable closed-loop dynamics. Assume that the three inner control loops in Fig. \ref{overall}, corresponding to the load racks, the HVAC, and the inverters, are subject to time delays $\tau_1$, $\tau_2$, and $\tau_3$, respectively. \newstuff{We treat time delays as constants on the timescale of the system dynamics, but they are allowed to vary on longer timescales.} The state-space model of the system shown in Fig. \ref{overall} takes the form
\begin{equation}\label{delaysys}
    \dot{x}(t) = A_0x(t) + \sum_{i=1}^3 A_{i} x(t-\tau_i),
  \vspace{-0.1cm}  
\end{equation}
where $x(t)\in \mathbf{R}^n$ and $A_i\in \mathbf{R}^{n\times n}$. The eigenvalues of the time-delay system (\ref{delaysys}) are given by the solution of the characteristic equation
$
    \text{det}\left( \lambda I-A_0-\sum_{i=1}^3 A_{i}e^{-\lambda \tau_i} \right) =0.
$
One way to approximate the eigenvalues of (\ref{delaysys}) is to interpolate the state history $x(\tau),\ \tau \in [-\tau_{\text{max}},0]$ with a Lagrange polynomial at $N+1$ points $-\tau_{\text{max}}=\theta_0\leq \cdots \leq \theta_{N}=0$. If the interpolation points are chosen to be the Chebyshev points $\theta_k = (\text{cos}(k\pi /N)-1)\tau_{\text{max}}/2$ such that $\theta_{k_i}\approx \tau_i$ with some $k_i$ for each $ i=1,2,3$, the eigenvalues of 
$
    A_N = 
    \begin{bmatrix}
    D_N \otimes I_n \\
    e_{N+1}^T\otimes A_0 +\sum_{i=1}^3 e_{k_i+1}^T \otimes A_{i}
    \end{bmatrix}
$
will approach the eigenvalues of (\ref{delaysys}) with $O(N^{-N})$ \cite{michiels2007stability}. Here, $e_i$ is an $(N+1)$-dimensional unit vector with the $i$th element being 1, and $D_N$ is the $nN\times n(N+1)$ Chebyshev differentiation matrix \cite{michiels2007stability}. Therefore, we can use the rightmost eigenvalue of $A_N$ as the stability index of the time-delay system (\ref{delaysys}) to guide the design of the PI controller parameters.

\begin{figure}[!t]
	\centering
	\includegraphics[width=3.2in]{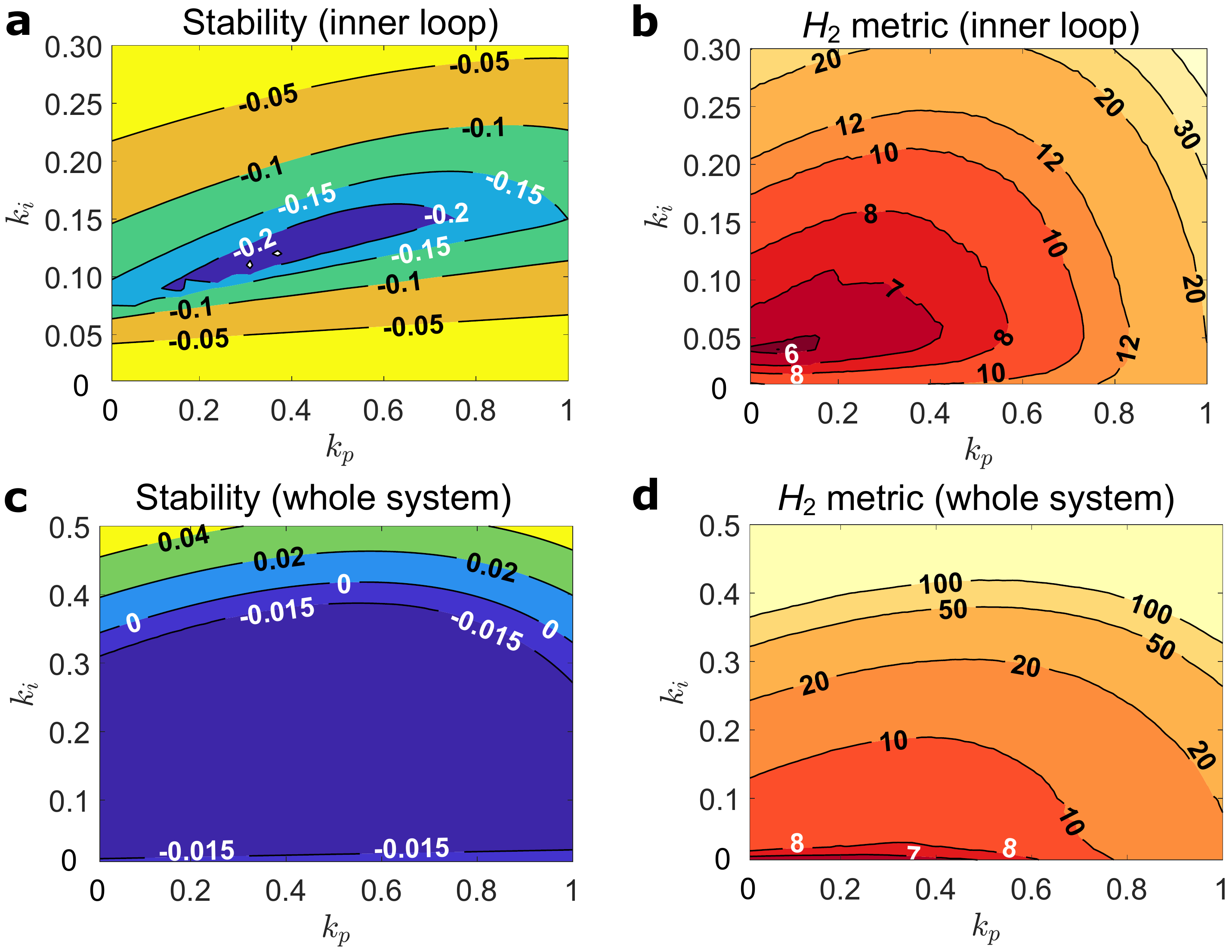}
	\caption{(a, c) Stability index and (b, d) performance metric of (a, b) the inner control loop for the load racks and (c, d) the whole system.} 
\label{contourdelay_kpki}
	\vspace{-0.4cm}   
\end{figure}

\vspace{-0.3cm}
\section{Hardware-in-the-Loop Tests}
\label{num}
Based on the methods discussed above, we design the PI controller parameters using the data obtained from the open-loop tests in Fig. \ref{LoadRack_Ind}. We then implement the proposed control structure in the HiL testbed and evaluate its performance.

\textit{Controller parameter design.}  The design of controller parameters is done in two stages. We first design the inner-loop FF PI controllers and then the outer-loop PI controller, based on the stability and performance analysis of the time-delay system. We take the Lyapunov exponent of the closed-loop system as its stability index and further quantify the control performance using an $H_{2}$ metric defined as
$\mathcal{M}  = \int_{0}^{t_m} (y(t)-y^*)^2 dt$
where $y(t)$ is system output, $y^*$ is the control target, and $t_m$ is a large enough time constant, which we choose to be $t_m=100s$. The three inner-loop FF PI controllers are designed independently. We take the one for the load racks as an example. Based on the open-loop tests in Fig. \ref{LoadRack_Ind}, we identify the response model for the load racks to be the first-order transfer function $G_1(s)=0.9359/(0.0890s+1)$ with a closed-loop time delay of $\tau_1=5.0 $s. The solution time of the device scheduler ($< 0.1$ s) is negligible compared with the communication delays. Figs. \ref{contourdelay_kpki}a-\ref{contourdelay_kpki}b show contour plots of the stability index and $H_{2}$ metric as functions of the FF PI parameters. Fig. \ref{contourdelay_kpki}a visualizes the region of stability and the point of maximum stability, and it shows that $k_p$ has a relatively small impact on the stability index while $k_i$ affects the stability index significantly. Combining Figs. \ref{contourdelay_kpki}a-\ref{contourdelay_kpki}b, we can design the controller parameters so that the system has good dynamical performance while maintaining a sufficient degree of stability; here we choose $k_p = 0.2$ and $k_i = 0.05$. The same analysis and design procedure are performed for the other two inner control loops. Having tuned all three lower-level FF PI controllers, we study the impact of the outer-loop PI controller on the overall system stability and performance. Figs. \ref{contourdelay_kpki}c-\ref{contourdelay_kpki}d show the contour plots for the stability index and $H_2$ metric of the whole system as functions of the outer-loop $k_p$ and $k_i$. This shows that, when $k_i$ is less than $0.3$, the system is stable and the stability index is insensitive to $k_p$. The best performance is achieved when both $k_p$ and $k_i$ are kept small, and here we choose $k_p = 0.15$ and $k_i = 0.05$.

% \begin{figure}[]
% 	\centering
% 	\includegraphics[width=1.7in]{delayKp.pdf}
% 	\includegraphics[width=1.7in]{delayKi.pdf}
% 	\caption{Simplified Control Loops} \label{lammax_vs_Kp}
% \end{figure}

% \begin{figure}[]
% 	\centering
% 	\includegraphics[width=3.2in]{contourdelay_full.png}
% 	\caption{Simplified Control Loops} \label{contourdelay_full}
% \end{figure}

% \begin{figure}[]
% 	\centering
% 	\includegraphics[width=3.2in]{contourH2_full.png}
% 	\caption{Simplified Control Loops} \label{contourH2_full}
% \end{figure}

% \begin{figure}[!ht]
% 	\centering
% 	\includegraphics[width=1.7in]{contourH2.pdf}
% 	\includegraphics[width=1.7in]{contourH2_full.pdf}
% 	\caption{Simplified Control Loops} \label{H2inner}
% \end{figure}

% \begin{figure*}[]
% 	\centering
% 	\includegraphics[width=6.0in]{LBNL_HIL_Paper_simplified.pdf}
% 	\caption{Configuration of the HIL Test} \label{testconfig}
% \end{figure*}

\begin{figure}[!t]
	\centering
	\includegraphics[width=3.2in]{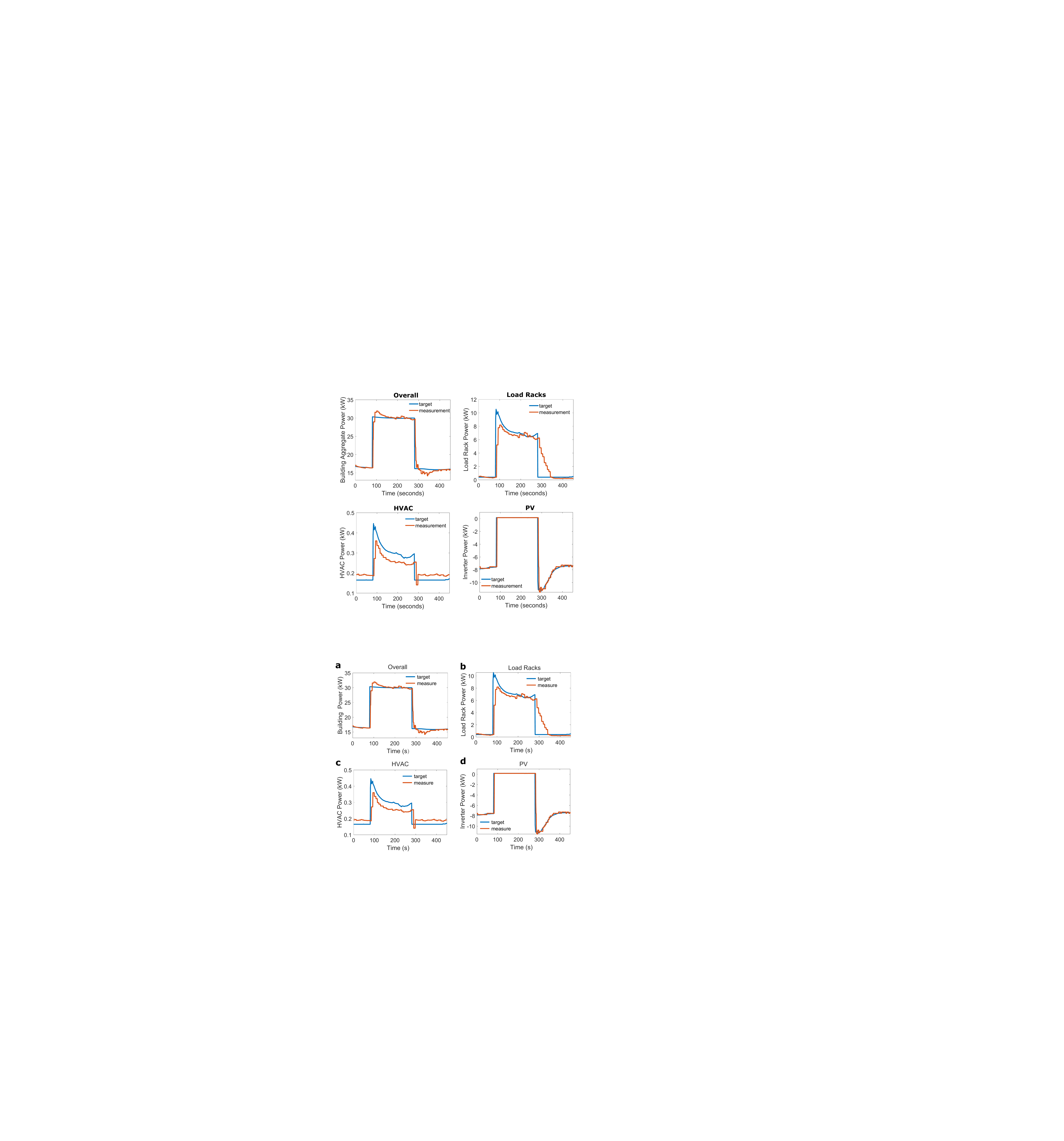}
\caption{Responses of the (a) overall building, (b) load racks, (c) HVAC, and (d) PV system.} 
	\label{overallpower_new}
	\vspace{-0.5cm}  
\end{figure}

\textit{Test results.} After implementing the control structure, we evaluate the building's overall response with two criteria, namely, the initial response time and the ramp time. The initial response time is the time it takes for the aggregate building power to start moving toward the target after the target is sent to the building-level controller. The ramp time is the time required for the aggregate building power to change from $0\%$ to $100\%$ of the target power. \newstuff{We send a square wave (blue curve in Fig. 5a) as the target for the building and measure the actual power (red curve in Fig. 5a) under the same device conditions as in the open-loop tests. } The individual responses of the load racks, HVAC, and PV system are shown in Figs. \ref{overallpower_new}b-\ref{overallpower_new}d, respectively. Compared to the open-loop control in Fig. \ref{LoadRack_Ind}, the proposed control method largely eliminates the steady-state control errors. Due to random variations in power demand in the building and random changes of the power quality, one can always observe small steady-state oscillations around the overall building target power of about 0.6kW in magnitude, which accounts for only 4.3\% of the magnitude of target change over the course of the experiment. The proposed control method significantly accelerates the overall response of the system. The observed initial response time and ramp time are around 1.8 s and 8.9 s, respectively. The proposed control structure ensures that the overall dynamical response of the building resembles that of the fast devices, whereas the flexibility from both the fast and slow devices are fully exploited in the steady state.

% \begin{figure}[ht!]
% 	\centering
% 	\includegraphics[width=1.7in]{overall_new.pdf}
% 	\caption{Aggregate building Power under the proposed control architecture.} \label{overallpower}
% \end{figure}

% \begin{figure}[!h]
% 	\centering
% 	\includegraphics[width=1.12in]{LoadRack_Res.pdf}
% 	 \includegraphics[width=1.12in]{HVAC_Res.pdf}
% 	 \includegraphics[width=1.12in]{Inverter_Res.pdf}
% 	 \caption{Target and observed power of different types of controllable devices under the proposed control architecture.}
%     \label{difftypes}
% \end{figure}

% \begin{figure}[ht!]
% 	\centering
% 	\includegraphics[width=2.8in]{inverterpower.png}
% 	\caption{Inverter (PV Battery System) Load Power} \label{inverterpower}
% \end{figure}

\vspace{-0.3cm}

\newstuff{
\section{Discussion}

\begin{figure}[t]
	\centering
	\includegraphics[width=3in]{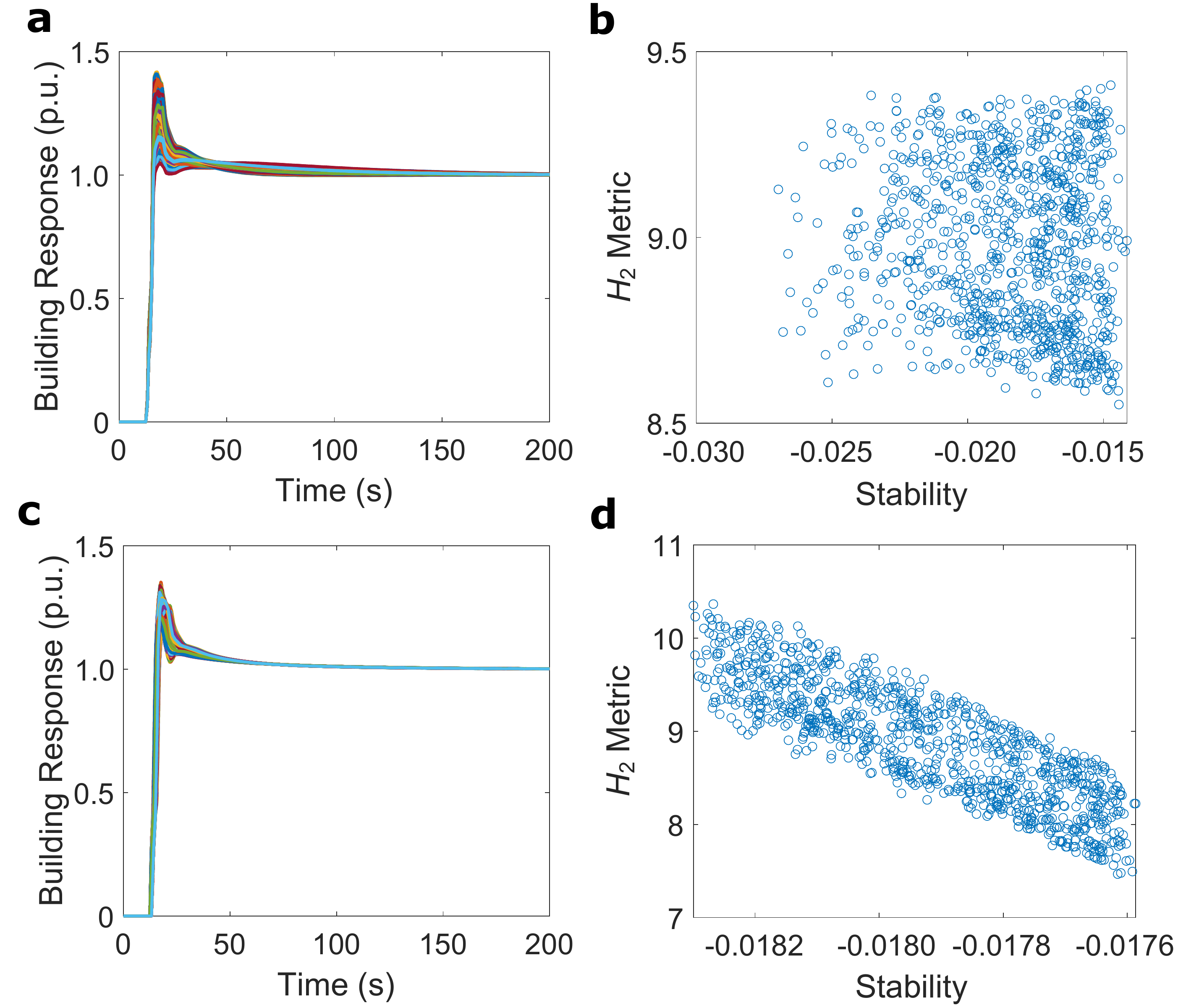}
\caption{\newstuff{(a, c) Time-domain responses and (b, d) performance metrics under $\pm 20\%$ uncertainties of the (a, b) steady-state gains and (c, d) time delays.}
} 
\label{impactuncertainties}
	\vspace{-0.3cm}  
\end{figure}

\begin{figure}[!h]
	\centering
	\includegraphics[width=3in]{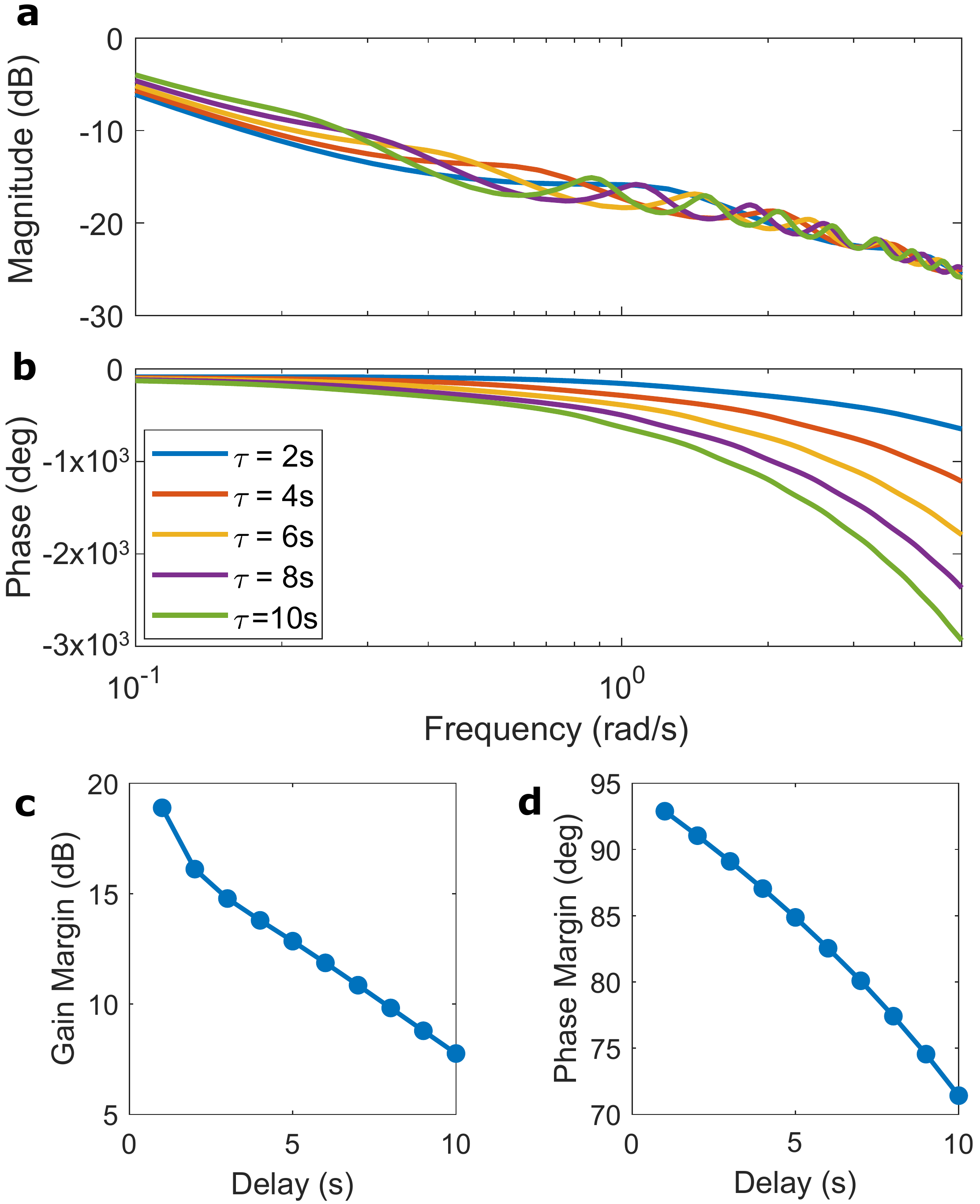}
\caption{\newstuff{(a) Magnitude and (b) phase plots of the Bode diagram under different time delays. (c) Gain and (d) phase margins as functions of the time delay.}} 
	\label{bodeplot}
	\vspace{-0.5cm}  
\end{figure}

\textit{Impact of uncertainties on responses.} The dynamical responses of the devices may be subject to uncertainties due to varying operation conditions. It is thus necessary to assess how uncertainties affect the stability and performance of the designed control system. Here, we perform simulation studies based on the models identified from the HiL tests. For each group of devices, we consider $\pm 20\%$ uncertainties in 1) the steady-state gain and 2) the time delay of each group of devices. For the two types of uncertainties considered, we perform simulations for 1000 realizations of the uncertain parameters. Figs. 6a-6b show the results under gain uncertainties, whereas Figs. 6c-6d illustrate the impact of delay uncertainties. The results show that $\pm 20\%$ uncertainties on the considered parameters have a notable impact on the dynamical responses (Figs. 6a and 6c) and both the stability index and the $H_2$ metric vary in certain ranges (Figs. 6b and 6d). The closed-loop system remains stable under all realizations of parameters, and thus the steady-state functionality remains intact. This shows a certain degree of robustness of the designed controller.

\textit{Impact of delays on stability margins.} The control system illustrated in Fig. 2 is formed by the open-loop system from tgt\_P to obs\_P with a unit negative feedback from obs\_P to err\_P. Therefore, we can study the relative stability of the closed-loop system using the Bode diagram of the open-loop system from tgt\_P to obs\_P. Here, we set the time delay of each inner loop to a common parameter $\tau$ and study the stability margins of the system as functions of $\tau$. Figs. 7a-7b show the magnitude and phase plots of the Bode diagram with different values of $\tau$. It is shown that a larger time delay leads to a larger phase shift at a given frequency and to reduced gain and phase crossover frequencies. This results in the deterioration of both gain and phase margins, as shown in Figs. 7c-7d. Though affected by the uncertainties, the designed controller maintains sufficient stability margins under a reasonably wide range of time delays.
}
\vspace{-0.3cm}

\section{Conclusion}
This experimental study reveals the negative impacts of time delays, \newstuff{uncertainties}, characterization errors, multiple timescales, and nonlinearity on the deployment of a DR program, and a set of control and design tools are introduced to address these practical issues. \newstuff{The proposed control method significantly reduces the initial response time from about 4.5 s to 1.8 s and also eliminates the steady-state errors, achieving a finite ramp time of 8.9 s. While unpredictable human behavior and the comfort level of building occupants have not been considered here, the designed controller is shown to be robustly stable even under $\pm 20\%$ parameter uncertainties. } We hope that this Letter will stimulate the consideration of practical challenges in future DR research.
\vspace{-0.3cm}

\bibliographystyle{ieeetr}

\end{document}